\documentclass[aps, prd, twocolumn,showkeys, showpacs,amsmath,amssymb,superscriptaddress]{revtex4}
\usepackage{graphicx}
\usepackage{dcolumn}
\usepackage{bm}
\usepackage{epsfig}
\usepackage{color}

\begin{document}
\title{Relativistic Lattice Boltzmann Model with Improved Dissipation}

\author{M. Mendoza} \email{mmendoza@ethz.ch} \affiliation{ ETH
  Z\"urich, Computational Physics for Engineering Materials, Institute
  for Building Materials, Schafmattstrasse 6, HIF, CH-8093 Z\"urich
  (Switzerland)}

\author{I. Karlin} \email{karlin@lav.mavt.ethz.ch} \affiliation{ETH
  Z\"urich, Department of Mechanical and Process Engineering,
  Sonneggstrasse 3, ML K 20, CH-8092 Z\"urich (Switzerland)}

\author{S. Succi} \email{succi@iac.cnr.it} \affiliation{Istituto per
  le Applicazioni del Calcolo C.N.R., Via dei Taurini, 19 00185, Rome
  (Italy),\\and Freiburg Institute for Advanced Studies,
  Albertstrasse, 19, D-79104, Freiburg, (Germany)}

\author{H. J. Herrmann}\email{hjherrmann@ethz.ch} \affiliation{ ETH
  Z\"urich, Computational Physics for Engineering Materials, Institute
  for Building Materials, Schafmattstrasse 6, HIF, CH-8093 Z\"urich
  (Switzerland)} \affiliation{Departamento de F\'isica, Universidade
  Federal do Cear\'a, Campus do Pici, 60455-760 Fortaleza, Cear\'a,
  (Brazil)}

\date{\today}
\begin{abstract}
  We develop a relativistic lattice Boltzmann (LB) model, providing a
  more accurate description of dissipative phenomena in relativistic
  hydrodynamics than previously available with existing LB schemes.
  The procedure applies to the ultra-relativistic regime, in which the
  kinetic energy (temperature) far exceeds the rest mass energy,
  although the extension to massive particles and/or low temperatures
  is conceptually straightforward. In order to improve the description
  of dissipative effects, the Maxwell-J\"uttner distribution is
  expanded in a basis of orthonormal polynomials, so as to correctly
  recover the third order moment of the distribution function.  In
  addition, a time dilatation is also applied, in order to preserve
  the compatibility of the scheme with a cartesian cubic lattice.  To
  the purpose of comparing the present LB model with previous ones,
  the time transformation is also applied to a lattice model which
  recovers terms up to second order, namely up to energy-momentum
  tensor.  The approach is validated through quantitative comparison
  between the second and third order schemes with BAMPS (the solution
  of the full relativistic Boltzmann equation), for moderately high
  viscosity and velocities, and also with previous LB models in the
  literature. Excellent agreement with BAMPS and more accurate results
  than previous relativistic lattice Boltzmann models are reported.
\end{abstract}


\maketitle

\section{Introduction}

Relativistic hydrodynamics and kinetic theory play a major role in
many forefronts of modern physics, from large-scale applications in
astrophysics and cosmology, to microscale electron flows in graphene
\cite{Geim1, natletter, grapPRL}, all the way down to quark-gluon
plasmas \cite{QGP-1, QGP-2, QGP-3}.  Due to their strong non-linearity
and, for the case of kinetic theory, high dimensionality as well, the
corresponding equations are extremely challenging even for the most
powerful numerical methods, let alone analytics. Recently, a promising
approach, based on a minimal form of relativistic Boltzmann equation,
whose dynamics takes place in a fully discrete phase-space and time
lattice, known as relativistic lattice Boltzmann (RLB), has been
proposed by Mendoza et al. \cite{rlbPRL, rlbPRD, LBE1}.  To date, the
RLB has been applied to the simulation of weakly and moderately
relativistic fluid dynamics, with Lorentz factors of $\gamma \sim
1.4$, where $\gamma = 1/\sqrt{1 - v^2/c^2}$, $c$ being the speed of
light and $v$ the speed of the fluid. This model reproduces correctly
shock waves in quark-gluon plasmas, showing excellent agreement with
the solution of the full Boltzmann equation as obtained by Bouras et
al. using BAMPS (Boltzmann Approach Multi-Parton Scattering)
\cite{BAMPS, BAMPSs}. The RLB makes use of two distribution functions,
the first one modeling the conservation of number of particles, and
the second one, the momentum-energy conservation equation.  The model
was constructed by matching the first and second order moments of the
discrete-velocity distribution function to those of the continuum
equilibrium distribution of a relativistic gas
%matching the first and second
%order moments of the equilibrium distribution function for
%relativistic particles.

In a subsequent work, Hupp et al.\cite{rlbhupp} improved the scheme by
extending the equilibrium distribution function for the number of
particles, in such a way as to include second order terms in the
velocity of the fluid, thereby taming numerical instabilities for
higher pressure gradients and velocities. However, the model was not
able to reproduce the right velocity and pressure profiles for the
Riemann problem in quark-gluon plasmas, for the case of large values
of the ratio between the shear viscosity and entropy density, $\eta/s
\sim 0.5$, at moderate fluid speeds ($v/c \sim 0.6$).

In order to set up a theoretical background for the lattice version of
the relativistic Boltzmann equation, Romatschke et al. \cite{PRCpaul}
developed a scheme for an ultrarelativistic gas based on the expansion
on orthogonal polynomials of the Maxwell-J\"uttner distribution
\cite{RelaBoltEqua} and, by following a Gauss-type quadrature
procedure, the discrete version of the distribution and the weighting
functions was calculated. This procedure was similar to the one used
for the non-relativistic lattice Boltzmann model \cite{HERMI, HERMI2}.
This relativistic model showed very good agreement with theoretical
data, although it was not compatible with a lattice, thereby requiring
linear interpolation in the free-streaming step.  This implies the
loss of some key properties of the standard lattice Boltzmann method,
such as negative numerical diffusion and exact streaming.

Very recently, Li et al. \cite{mrtrlbPRD} noticed that the equation of
conservation for the number of particles, recovered by the RLB model
\cite{rlbPRL, rlbPRD}, exhibits incorrect diffusive
effects. Therefore, they proposed an improved version of RLB, using a
multi-relaxation time collision operator in the Boltzmann equation,
showing that this fixes the issue with the equation for the
conservation of the number of particles.  The generalized collision
operator allows to tune independently the bulk and shear viscosities,
yielding results for the Riemann problem closer to BAMPS \cite{BAMPS}
when the bulk viscosity is decreased. However, the third order moment
of the equilibrium distribution still does not match its continuum
counterpart and therefore the model still has problems to reproduce
high $\eta/s \sim 0.5$, for moderately high velocities, $\beta = v/c =
0.6$.  Thus, while surely providing an improvement on the original RLB
model, the work \cite{mrtrlbPRD} did not succeed in reproducing the
vanishing bulk viscosity, which is pertinent to the ultra-relativistic
gas, while allowing the bulk viscosity to vary independently on the
shear viscosity.

Note that in the much more studied case of the lattice Boltzmann
models for the non-relativistic fluids, the question of the choice of
the lattice with higher-order symmetry requirements has only recently
been solved, in the framework of the entropy-compliant constriction
\cite{chikatamarla2006,chikatamarla2009}. However, the lattices
(space-filling discrete-velocity sets) found in that case are tailored
to reproduce the moments of the non-relativistic Maxwell-Boltzmann
distribution, and do not seem to be directly transferable to the
present case of the relativistic (Maxwell-J\"uttner) equilibrium
distribution, which has fairly different symmetries as compared to the
non-relativistic Maxwell-Boltzmann distribution.  Therefore, the
extension of the previous LB models has to be considered anew.

In this paper, we develop a new lattice Boltzmann model capable of
reproducing the third order moment of the continuum equilibrium
distribution, and still realizable on a cubic lattice.  The model is
based on a single distribution function and satisfies conservation of
both number of particles and momentum-energy equations.  The model is
based on the single relaxation time collision operator proposed by
Anderson and Witting \cite{Anderson, RelaBoltEqua} which is more
appropriate for the ultra-relativistic regime than the Marle model
used in the previous works, Thus, the proposed model offers
significant improvement on previous relativistic lattice Boltzmann
models in two respects: (i) It captures the symmetry of the
higher-order equilibrium moments sufficiently to reproduce the
dissipative relativistic hydrodynamics at the level of the Grad
approximation to the relativistic Boltzmann equation; (ii) It
represents a genuine lattice Boltzmann discretization of space and
time, with no need of any interpolation scheme, thereby avoiding the
otherwise ubiquitous spurious dissipation. The new lattice Boltzmann
model is shown to reproduce with very good accuracy the results of the
shock-waves in quark-gluon plasmas, for moderately high velocities and
high ratios $\eta/s$.

The paper is organized as follows: in Sec. \ref{sec:model} we describe
in detail the model and the way it is constructed; in
Sec.~\ref{sec:num}, we implement simulations of the Riemann problem in
order to validate our model and compare it with BAMPS and previous
relativistic lattice Boltzmann models; finally, in
Sec.~\ref{sec:diss}, we discuss the results and future work.

\section{Model Description}\label{sec:model}

\subsection{Symmetries of the relativistic Boltzmann equation}

To build our model, we start from the relativistic Boltzmann equation
for the probability distribution function $f$:
\begin{equation}\label{Boltzmann:eq}
  p^\mu \partial_\mu f = -\frac{p_\mu U^\mu}{c^2 \tau}( f - f^{\rm eq} ) \quad ,
\end{equation}
where the local equilibrium is given by the Maxwell-J\"uttner
equilibrium distribution
\cite{RelaBoltEqua}, \begin{equation}\label{MJ:eq} f^{\rm eq} = A
  \exp(-p^\mu U_\mu/k_B T) \quad ,
\end{equation}
In the above, $A$ is a normalization constant, $c$ the speed of light,
and $k_B$ the Boltzmann constant.  The $4$-momentum vectors are
denoted by $p^\mu = (E/c, \vec{p})$, and the macroscopic 4-velocity by
$U^\mu = (c, \vec{u}) \gamma (u)$, with $\vec{u}$ the
three-dimensional velocity of the fluid. Note that we have used the
Anderson-Witting collision operator\cite{Anderson} (rhs of
Eq. \eqref{Boltzmann:eq}), making our model compatible with the
ultrarelativistic regime.  Hereafter, we will use natural units, $c =
k_B = 1$, and work in the ultrarelativistic regime, $\xi \equiv
mc^2/k_B T \ll 1$.

According to a standard procedure \cite{HERMI, HERMI2, PRCpaul}, we
first expand the Maxwell-J\"uttner distribution in the rest frame,
$f^{\rm eq} = A \exp(-p^0/T)$, in an orthogonal basis.  Since in the
ultrarelativistic regime, $p^0/T = \sqrt{\vec{p}^2/T^2 + m^2/T^2}
\simeq p/T$, being $p = \sqrt{\vec{p}^2}$, we can write the
equilibrium distribution in spherical coordinates,
%and treat the
%contributions of each coordinate independently,
% \begin{equation}\label{eq:spherical}
%   \int e^{-p_0/T} \frac{d^3p}{p^0} = \int_0^\infty p
%   e^{-p/T} dp \int_0^\pi \sin(\theta) d\theta \int_0^{2\pi} d\phi
%   \quad .
% \end{equation}
\begin{equation}
  \int g e^{-p_0/T} \frac{d^3p}{p^0} = \int_0^\infty\int_0^\pi \int_0^{2\pi} g p
  e^{-p/T} dp \sin(\theta) d\theta d\phi \quad , 
\end{equation}
where $g$ is an arbitrary function of momentum. Following Romatschke
\cite{PRCpaul}, we can expand the distribution in each coordinate
separately, and subsequently, by using a Gauss quadrature, calculate
the discrete values of the 4-momentum vectors. Thus, the discrete
equilibrium distribution can be written as,
\begin{equation}
  f_l^{\rm eq} = \sum_{i,j,k} a_{ijk}(U^\mu) P_i(\theta_l) {\cal R}_j(p_l) F_k(\phi_l) \quad ,
\end{equation}
where the coefficients $a_{ijk}(U^\mu)$ are the projections of the
distribution on the polynomials $P_i(\theta_l) {\cal R}_j(p_l)
F_k(\phi_l)$, and the discrete 4-momenta are denoted by $p_l^\mu=(p_l,
p_l \cos(\phi_l) \sin(\theta_l), p_l \sin(\phi_l) \sin(\theta_l), p_l
\cos(\theta_l) )$. Consequently, the discrete form of the Boltzmann
equation takes the form,
\begin{equation}\label{discRLB:eq}
  f_l(x^\mu + p_l^\mu/p_l^0 \delta t, t + \delta t) - f_l(x^\mu, t) =
  -\frac{p_{l\mu} U^\mu \delta t}{\tau p_l^0}( f_l - f_l^{\rm eq} ) \quad .
\end{equation}
However, note that, in the streaming process on the right-hand-side of
Eq.\eqref{discRLB:eq}, the distribution moves at velocity
$p_l^\mu/p_l^0$, which implies that the information travels (in a
single time step) from each cell center to a position that belongs to
the surface of a sphere of radius $c\;\delta t = 1$.  Furthermore, to
represent correctly the third order moment of the equilibrium
distribution,
\begin{equation}
 P^{\alpha \beta \lambda} = \sum_l f^{\rm eq}_l p_l^\alpha p_l^\beta p_l^\lambda \quad ,
\end{equation}
the number of points needed on the surface of the unit sphere exceeds
$6$ and $12$, which correspond to the first neighbors for a cubic and
hexagonal closed packed (HCP) lattices, respectively.  This implies
that, in general, the 4-vectors $p^\mu/p^0$ cannot be embedded into a
regular lattice, and therefore, an interpolation algorithm has to be
used. By doing this, we are losing one of the most important features
of lattice Boltzmann models, which is the exact streaming.  Thus,
within this spherical coordinate representation, the streaming process
cannot take place on a regular lattice.

\subsection{Moment projection of the equilibrium}

In this work, we shall use a different approach to the quadrature
representation.  We first expand the equilibrium distribution at rest,
$w(p^0) = f^{\rm eq} = A \exp(-p^0)$ by using Cartesian coordinates,
unlike the spherical coordinate system used in Ref.~\cite{PRCpaul},
and choose the 4-momentum vectors such that they belong to the lattice
(from now on and without loss of generality, we will use the notation
$p^0 /T \rightarrow p^0$).  This procedure also avoids extra terms in
the product, $P_i(\theta_l) {\cal R}_j(p_l) F_k(\phi_l)$ for the
spherical case, which are not necessary if we only need to recover
correctly the first three moments of the equilibrium distribution.
This considerably simplifies the discrete equilibrium distribution.

By performing a Gram-Schmidt procedure with the weight $w(p^0)$, we
construct a set of orthonormal polynomials.  The orthonormal
polynomials in cartesian coordinates up to third order, herefrom
denoted by $J_k$, where the index $k$ runs from $0$ to $29$, are shown
in Table \ref{table1}.
\begin{table}
  \centering
  \begin{tabular}{|c|c|c|}\hline
    Order & Polynomial $J_k$ & $k$\\ \hline
    $0$th & 1  & 0 \\ \hline
    $1$st & $\frac{ p^0-2}{\sqrt{2}}$,  $\frac{ p^x}{\sqrt{2}}$,
    $\frac{p^y}{\sqrt{2}}$, $\frac{p^z}{\sqrt{2}}$ & 1, 2, 3, 4\\ \hline
    $2$nd &  $\frac{( p^0 -6) p^0+6}{2
      \sqrt{3}}$, $\frac{(p^0-4) p^x}{2 \sqrt{2}}$,  $\frac{(p^0-4) p^y}{2 \sqrt{2}}$ & 5, 6, 7 \\
    & $\frac{(p^0-4) p^z}{2 \sqrt{2}}$, $\frac{-p^{0 2}+p^{x 2}+2 p^{y 2}}{4 \sqrt{2}}$, $-\frac{p^{0 2}-3
      p^{x 2}}{4 \sqrt{6}}$ &
    8, 9, 10 \\
    & $\frac{p^x p^z}{2 \sqrt{2}}$, $\frac{p^y p^z}{2 \sqrt{2}}$, $\frac{p^x p^y}{2
      \sqrt{2}}$
    & 11, 12, 13\\\hline
    $3$rd & $\frac{1}{12} ( p^0-6)^2 p^0-2$,  $ \frac{((p^0-10) p^0+20)
      p^x}{4 \sqrt{5}}$  & 14, 15 \\
    & $-\frac{1}{24} (p^0-6)\left(p^{0 2}-3 p^{x 2}\right)$, $\frac{5
      p^{x 3}-3 p^{0 2} p^x}{24 \sqrt{5}}$ & 16, 17 \\
    & $\frac{((p^0-10) p^0+20) p^y}{4 \sqrt{5}}$, $\frac{(p^0-6) p^x
      p^y}{4 \sqrt{3}}$, $\frac{p^x p^y p^z}{4 \sqrt{3}}$ & 18, 19, 20\\
    &  $-\frac{(p^0-6) \left(p^{0 2}-p^{x 2}-2 p^{y 2}\right)}{8
      \sqrt{3}}$, $\frac{p^x \left(-p^{0 2}+p^{x 2}+2 p^{y
          2}\right)}{8 \sqrt{3}}$ & 21, 22 \\
    & $\frac{p^y \left(-3 p^{0 2}+3 p^{x 2}+4 p^{y 2}\right)}{24
      \sqrt{2}}$, $\frac{((p^0-10) p^0+20) p^z}{4 \sqrt{5}}$ & 23, 24
    \\ & $\frac{(p^0-6) p^x p^z}{4 \sqrt{3}}$, $-\frac{p^z \left(p^{0
          2}-5 p^{x 2}\right)}{8 \sqrt{30}}$, $\frac{(p^0-6) p^y
      p^z}{4 \sqrt{3}}$ & 25, 26, 27 \\
    & $\frac{(p^0-6) p^y p^z}{4 \sqrt{3}}$, $\frac{p^z \left(-p^{0 2}+p^{x 2}+4
        p^{y 2}\right)}{24 \sqrt{2}}$ & 28, 29 \\ \hline
  \end{tabular}
  \caption{Polynomials $J_k$ that are orthonormal on the weight
    function $w(p^0)$ in Cartesian coordinates $(x,y,z)$.}
  \label{table1}
\end{table}
Note that in this Table, the 4-momentum has the notation $p^\mu =
(p^0, p^x, p^y, p^z)$. Since these polynomials are orthonormal, there
are no normalization factors, and the Maxwell-J\"uttner distribution
can be approximated, up to third order in the momentum space, by an
expansion as simple as
\begin{equation}\label{eq:expanded}
  f^{\rm eq} \simeq \sum_{k=0}^{29} w(p^0) a_{k}(T, U^\mu) J_k(p^\mu) \quad ,
\end{equation}
where the projections $a_k$ are calculated by,
\begin{equation}\label{coeff}
  a_k = \int f^{\rm eq} J_k(p^\mu) \frac{d^3p}{p^0} \quad .
\end{equation}

Since the Anderson-Witting model is only compatible with the
Landau-Lifshitz decomposition \cite{RelaBoltEqua, Anderson}, we must
calculate the energy density of the fluid by solving the eigenvalue
problem,
\begin{equation}\label{eigenproblem}
  T^{\alpha \beta} U_\beta = \epsilon U^\alpha \quad ,
\end{equation}
$\epsilon$ being the energy density of the fluid, and
\begin{equation}
  T^{\alpha \beta} = \int f  p^\alpha p^\beta \frac{d^3p}{p^0}\quad ,
\end{equation}
the momentum-energy tensor.  For the particle density, we use the
relation,
\begin{equation}\label{densitycal}
  n = U_\alpha \int f p^\alpha \frac{d^3p}{p^0} \quad ,
\end{equation}
and, by using the equation of state, $\epsilon = 3 n T$, we can
calculate the temperature of the fluid.

\subsection{Discrete-velocity representation of the quadratures}

Note that the above derivation using Cartesian coordinates still
refers to the continuous 4-momenta.  In order to discretize the above
moment projection of the equilibrium distribution, we must choose a
set of 4-momentum vectors that satisfies the very same orthonormality
conditions, namely:
\begin{equation}\label{ortho:qua}
  \int w(p^0) J_l(p^\mu) J_k(p^\mu) \frac{d^3p}{p^0} = \sum_i w_i J_l(p^\mu_i)
  J_k(p^\mu_i) = \delta_{lk} \quad ,
\end{equation}
while, at the same time, $p^\mu/p^0$ corresponds to lattice
points. Here, we choose to work with a cubic lattice, although the
procedure described here also applies to other ones, e.g. HCP lattice.

Since, due to its nature, $p^\mu/p^0$ leads to velocity vectors which
belong to a sphere of radius $c$ in the space components, using the
procedure in Ref. \cite{PRCpaul} will generally result in off-site
lattice points.  For this reason, we opt for another quadrature based
on this orthonormality condition, and impose that the distribution
function at rest frame should satisfy the moments of the equilibrium
distribution, up to $6$-th order. This is made to ensure that the
$5$-th order moment of the equilibrium distribution is recovered (at
least at very low fluid velocities), which, in the context of the Grad
theory for the Anderson-Witting model \cite{RelaBoltEqua}, is a
requirement for the correct calculation of the transport coefficients,
namely the shear and bulk viscosities and thermal conductivity.  The
condition for the $6$-th order moment, is to choose from the multiple
lattice solutions, the one that presents the highest symmetry to model
the Maxwell-J\"uttner distribution. In order to use general features
of classical lattice Boltzmann models, like bounce-back boundary
conditions to impose zero velocity on solid walls, we will also
require that the weights $w_i$ corresponding to the discrete
4-momentum vectors $p_i^k$ have the same values as the ones
corresponding to $-p_i^k$ (latin indices run over spatial components).

\begin{figure}
\centering
\includegraphics[scale=0.8]{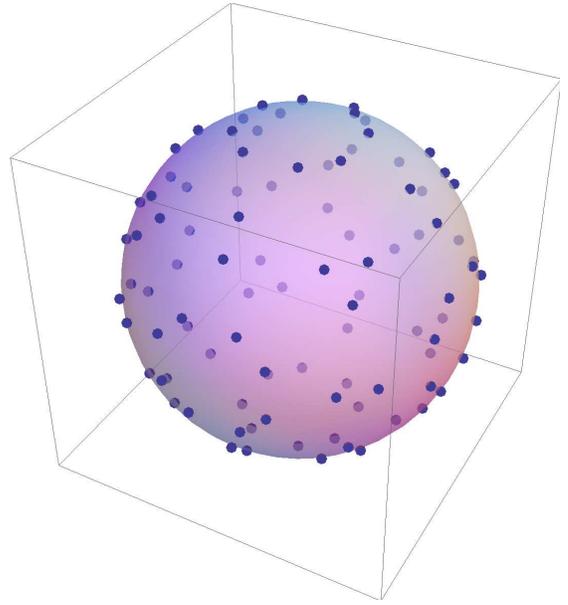}
\caption{Directions of the velocity vectors $\vec{\vartheta}_i$ to
  recover up to the third order moment of the Maxwell-J\"uttner
  distribution. The radius of the sphere is $R=\sqrt{41}$. The points
  represent lattice sites belonging to the sphere surface.}
\label{fig1}
\end{figure}

In order to generate on-site lattice points, let us first analyse the
relativistic Boltzmann equation, which can be written as,
\begin{equation}\label{Boltzmann2:eq}
  p^0 \partial_t f + p^a \partial_a f = -\frac{p_\mu U^\mu}{\tau}( f - f^{\rm eq} ) \quad ,
\end{equation}
and in the ultrarelativistic regime,
\begin{equation}\label{Boltzmann3:eq}
  p^0 (\partial_t f + v^a \partial_a f) = -\frac{p_\mu U^\mu}{\tau}( f - f^{\rm eq} ) \quad .
\end{equation}
where $v^a$ are the components of the microscopic velocity.  These
microscopic velocities have the same magnitude but, in general,
different directions. Dividing both sides of Eq.\eqref{Boltzmann3:eq}
by $p^0$, we obtain
\begin{equation}\label{Boltzmann3:eq}
  \partial_t f + v^a \partial_a f = -\frac{p_\mu U^\mu}{ \tau p^0}( f - f^{\rm eq} ) \quad .
\end{equation}
In other words, in the ultra-relativistic regime, the relativistic
Boltzmann equation can be cast into a form where the time derivative
and the propagation term become the same as in the non-relativistic
case, at the price of an additional dependence on $p^0$ in the
relaxation term.  However, since this newly acquired dependence
remains local, we shall be able to find a discrete-velocity quadrature
which also allows for a lattice Boltzmann-type discretization in time
and space without any interpolation. Indeed, in a cubic cell of length
$\delta x = 1$ there are only $6$ neighbors, which are not sufficient
to satisfy the orthogonality conditions and the third order moment of
the equilibrium distribution.  However, by multiplying this equation
by a constant $R$ at both sides, and performing a time transformation
(dilatation), $\delta t \rightarrow R \delta t'$ and $\tau \rightarrow
R \tau'$, we obtain
\begin{equation}\label{Boltzmann3:eq}
  \partial_{t'} f + \vartheta^a \partial_a f = -\frac{p_\mu U^\mu}{ \tau' p^0}( f - f^{\rm eq} ) \quad ,
\end{equation}
where we have defined $\vartheta^a = R v^a$. Due to this
transformation, the 4-momentum vectors are reconstructed through the
relation
\begin{equation}\label{eq:mrecon}
  p^\mu = p^0 ( 1, \vec{\vartheta}/R) \quad ,
\end{equation}
At this stage, we can choose the radius of the sphere such that the
lattice points that belong to the surface of the sphere and the cubic
lattice exhibit enough symmetries to satisfy both conditions.  This is
equivalent to solving the Diophantine equation,
\begin{equation}\label{eq:dio}
  n_x^2 + n_y^2 + n_z^2 = R^2 \quad ,
\end{equation}
where $n_x$, $n_y$, and $n_z$ are integer numbers, being
$\vec{\vartheta} = (n_x, n_y, n_z)$. Thus, we can determine the
components of the discrete version of the velocities $\vec{\vartheta}$
which are needed for the streaming term in the Boltzmann equation, lhs
of Eq.~\eqref{Boltzmann3:eq}.  However, on the rhs of this equation,
and for the calculation of the discrete 4-momentum vectors via
Eq.~\eqref{eq:mrecon}, we also need to know the discrete values of
$p^0$. The 4-vector $p^\mu$ is needed to compute the orthonormality
conditions given by Eq.~\eqref{ortho:qua} and the moments of the
equilibrium distribution.

Due to the fact that $p^0$ is the magnitude of the 4-momentum, $p^0 =
\sqrt{p^\mu p_\mu}$, in $3+1$-dimensional spacetime, it is natural to
assume that its discrete values can be calculated by using the weight
function in spherical coordinates, $w(p) = 4 \pi A p^2 \exp(-p)$,
where the angular components have been integrated out, and using the
zeros of its respective orthonormal polynomial of fourth order (this
is because we are interested in an expansion up to third order, so we
need one more order to calculate the zeros). This fourth order
polynomial is given by:
\begin{equation}
  {\cal R}^{(4)}(p) = \frac{1}{24 \sqrt{5}}[120 + p(-240 + p[120 + (p-20)p])]
  \quad .
\end{equation}

To summarize, in order to calculate the discrete $p^\mu_i$ and their
respective $w_i$, we first fix $R$ and solve the equations
\begin{subequations}
  \begin{equation}
    n_x^2 + n_y^2 + n_z^2 = R^2 \quad ,
  \end{equation}
  \begin{equation}
     {\cal R}^{(4)}(p) = 0 \quad ,
   \end{equation}
\end{subequations}
to obtain the solutions for $n_x$, $n_y$, $n_z$, and $p$.  With these
values, we build the discrete 4-vectors
\begin{equation}\label{eq:mrecon2}
  p_{lm}^\mu = p_l^0 ( 1, n_{x, m}/R, n_{y, m}/R, n_{z, m}/R) \quad ,
\end{equation}
where $l = 1, ..., 4$ denotes the four zeros of the polynomial ${\cal
  R}^{(4)}(p)$, and $m=0, ..., {\cal M}$ the triplets $(n_x, n_y,
n_z)_m$ that satisfy the Diophantine equation, assuming that ${\cal
  M}$ is the number of solutions. Here, for simplicity, we regroup the
pair of indexes $_{lm}$ to $_i$, so that we can label the discrete
4-momentums as $p^\mu_i$, where $i = 1, ..., {\cal N}$ with ${\cal N}
= 4\times {\cal M}$.

Next, we replace these values into the equations,
\begin{subequations}\label{eqs:eqs}
  \begin{equation}
    \int w(p^0) J_l(p^\mu) J_k(p^\mu) \frac{d^3p}{p^0} = \sum_i^{\cal N} w_i J_l(p^\mu_i)
    J_k(p^\mu_i) = \delta_{lk} \quad ,
  \end{equation}
  \begin{equation}
    \int w(p^0) p^\mu p^\nu p^\sigma p^\lambda \frac{d^3p}{p^0} =
    \sum_i^{\cal N} w_i p_i^\mu p_i^\nu p_i^\sigma p_i^\lambda \quad ,
  \end{equation}
  \begin{equation}
    \int w(p^0) p^\mu p^\nu p^\sigma p^\lambda p^\gamma \frac{d^3p}{p^0} =
    \sum_i^{\cal N} w_i p_i^\mu p_i^\nu p_i^\sigma p_i^\lambda p_i^\gamma \quad ,
  \end{equation}
  \begin{equation}
    \int w(p^0) p^\mu p^\nu p^\sigma p^\lambda p^\gamma p^\beta \frac{d^3p}{p^0} =
    \sum_i^{\cal N} w_i p_i^\mu p_i^\nu p_i^\sigma p_i^\lambda p_i^\gamma p_i^\beta \quad ,
  \end{equation}
  \begin{equation}
    w_i = w_j \quad \text{(if $p^k_i = - p^k_j$)} \quad ,
  \end{equation}
  \begin{equation}
    w_i \geq 0 \quad ,
  \end{equation}
\end{subequations}
and look for any solution for $w_i$ that fulfills the above relations.
Should none be found, we repeat the procedure with a different value
of $R$.  By performing this iteration process, we found that
$R=\sqrt{41}$ is sufficient to recover up to the third order moment of
the Maxwell-J\"uttner distribution, and up to sixth order of this
distribution in the Lorentz rest frame.

The corresponding discrete velocity vectors $\vec{\vartheta}_m$ are:
$(\pm 6, \pm 2, \pm 1)$, $(\pm 6, \pm 1, \pm 2)$, $(\pm 2, \pm 6, \pm
1)$, $(\pm 1, \pm 6, \pm 2)$, $(\pm 1, \pm 2, \pm 6)$, $(\pm 2, \pm 1,
\pm 6)$, $(\pm 5, 0, \pm 4)$, $(\pm 5, \pm 4, 0)$, $(0, \pm 5, \pm
4)$, $(\pm 4, \pm 5, 0)$, $(0, \pm 4, \pm 5)$, $(\pm 4, 0, \pm 5)$,
$(\pm 4, \pm 3, \pm 4)$, $(\pm 3, \pm 4, \pm 4)$, and $(\pm 4, \pm 4,
\pm 3)$; with the values for $p_l^0 \simeq 0.743$, $2.572$, $5.731$,
and $10.95$. Consequently, this gives a total of 4-momentum vectors
${\cal N} = 384$. However, the last condition in Eq.~\eqref{eqs:eqs}
allows some weights to become zero. Therefore, in our iteration
procedure, we have taken the minimal number of 4-momentum vectors
$p^\mu_i$, by imposing the maximum number of $w_i$ to be zero. For
this reason, there are only $128$ vectors $p^\mu_i$ needed to fulfill
the conditions in Eq.~\eqref{eqs:eqs}. In principle, all the velocity
vectors $\vec{\vartheta}_m$ are needed, but only some of the
combinations with $p^0_l$ are required. The detailed list of the
$\vec{\vartheta}_m$, $p^0_l$, and $p^\mu_i$, and their respective
discrete weight functions $w_i$ are given in the Supplementary
Material \cite{supp}.

In Fig.~\ref{fig1} we report the configuration of the velocity vectors
$\vec{\vartheta}$ to achieve the third order moment of the
Maxwell-J\"uttner distribution function. The points correspond to
lattice nodes of a cubic lattice that, at the same time, belong to the
surface of the respective sphere of radius $R = \sqrt{41}$.  The
relatively large number of discrete velocities should not come as a
surprise; in the case of non-relativistic lattice Boltzmann, the
number of discrete velocities also becomes high (at least 41 for
achieving complete Galilean invariance in the non-thermal case and
$125$ in the thermal case, see
\cite{chikatamarla2006,chikatamarla2009}).  Note that the specified
values of $p^0$ play the same role in defining the quadrature as the
reference temperature (energy) in the non-relativistic case
\cite{chikatamarla2006,chikatamarla2009}.

Finally, we can write the discrete version of the equilibrium
distribution up to third order,
\begin{equation}\label{eq:disc1}
  f_i^{\rm eq} = w_i \sum_{n=0}^{29} a_{n}(T, U^\mu) J_n(p_i^\mu)
  \quad ,
\end{equation}
which is shown in details in Appendix \ref{eqfuncapp},
Eq.~\eqref{eq:equi3rd}.  Note that this distribution function recovers
the first three moments of the Maxwell-J\"uttner distribution in the
ultrarelativistic regime,
\begin{equation}\label{moments:dis1}
  \int f^{\rm eq} p^\mu \frac{d^3p}{p^0} = \sum_{i=1}^{128} f_i^{\rm eq} p_i^\mu = N^\mu \quad ,
\end{equation}
\begin{equation}\label{moments:dis2}
  \int f^{\rm eq} p^\mu p^\nu \frac{d^3p}{p^0} = \sum_{i=1}^{128} f_i^{\rm eq} p_i^\mu p_i^\nu = T^{\mu \nu} \quad ,
\end{equation}
\begin{equation}\label{moments:dis3}
  \int f^{\rm eq} p^\mu p^\nu p^\lambda \frac{d^3p}{p^0} = \sum_{i=1}^{128} f_i^{\rm eq} p_i^\mu p_i^\nu p^\lambda = P^{\mu \nu \lambda} \quad ,
\end{equation}
where
\begin{equation}
  N^\nu = n U^\nu \quad ,
\end{equation}
\begin{equation}
  T^{\nu \mu} = -nT \eta^{\nu \mu} + 4nT U^{\nu} U^{\mu} \quad ,
\end{equation}
being the number of particles 4-flow and the energy-momentum tensor,
respectively, and
\begin{equation}
  \begin{aligned}
    P^{\nu \mu \lambda} = - 4nT^2 (\eta^{\nu \mu}U^{\lambda} + \eta^{\nu
      \lambda} U^{\mu} &+ \eta^{\mu \lambda} U^{\nu}) \\ &+ 24 nT^2 U^\nu
    U^\mu U^\lambda \quad ,
  \end{aligned}
\end{equation}
with $n = 2 T^3$. However, the extension to the case of massive
particles is straightforward, by changing the coefficients, $a_n$, in
Eqs.~\eqref{coeff} and \eqref{eq:disc1}.

\subsection{Discrete relativistic Boltzmann equation}

In the model of Anderson-Witting for the collision operator, the
relativistic Boltzmann equation takes the form given by
Eq.~\eqref{Boltzmann:eq},
\begin{equation}\label{Boltzmann:eq2}
  p^\mu \partial_\mu f = -\frac{p^\mu U_\mu}{\tau}( f - f^{\rm eq} ) \quad .
\end{equation}
This collision operator is compatible with the Landau-Lifshitz
decomposition \cite{RelaBoltEqua}, which implies fulfillment of the
following relations
\begin{subequations}
  \begin{equation}
    U_\mu N^\mu = U_\mu \int f p^\mu \frac{d^3p}{p^0} = U_\mu N_E^\mu
    = \int f^{\rm eq} p^\mu \frac{d^3p}{p^0} \quad ,
  \end{equation}
  \begin{equation}
    U_\mu T^{\mu \nu} = U_\mu \int f p^\mu p^\nu \frac{d^3p}{p^0} =
    U_\mu T_E^{\mu \nu} = \int f^{\rm eq} p^\mu p^\nu \frac{d^3p}{p^0} \quad ,
  \end{equation}
\end{subequations}
Here, the subscript $E$ denotes the quantities calculated with the
equilibrium distribution. Therefore, upon integrating
Eq.~\eqref{Boltzmann:eq2} in momentum space, we obtain
\begin{equation}\label{eq:conser1}
  \partial_\mu N^\mu = 0 \quad ,
\end{equation}
which is the conservation of the number of particles 4-flow.  By
multiplying by $p^\nu$ and integrating, we obtain the conservation of
the momentum energy tensor
\begin{equation}\label{eq:conser2}
  \partial_\mu T^{\mu \nu} = 0 \quad .
\end{equation}
In order to calculate the transport coefficients, we need the third
order moment, so that, upon multiplying Eq.~\eqref{Boltzmann:eq2} by
$p^\nu p^\beta$, we obtain
\begin{equation}\label{eq:conser3}
  \partial_\mu P^{\mu \nu \beta} = -\frac{1}{\tau} (U_\mu P^{\mu \nu \beta} - U_\mu P_E^{\mu \nu \beta}) \quad ,
\end{equation}
and by using a Maxwellian iteration method \cite{RelaBoltEqua},
\begin{equation}\label{eq:conser4}
  U_\mu P^{\mu \nu \beta} - U_\mu P_E^{\mu \nu \beta} = -\tau \partial_\mu P_E^{\mu \nu \beta} \quad .
\end{equation}
Note that we need at least the third order moment of the equilibrium
distribution, $P_E^{\mu \nu \beta}$, to compute the dissipation
coefficients (namely, bulk and shear viscosities and heat
conductivity). This requirement is fulfilled in our discrete and
continuum expansions of the equilibrium distribution via
Eqs.~\eqref{moments:dis1}, \eqref{moments:dis2},
\eqref{moments:dis3}. However, to recover full dissipation, we would
also need to recover the third moment of the non-equilibrium
distribution, which according to the $14$ moments Grad's theory , can
be written as,
\begin{equation}
  P^{\mu \nu \beta} = P_E^{\mu \nu \beta} + b_\alpha P_E^{\mu \nu
    \beta \alpha} + d_{\alpha \lambda} P_E^{\mu
    \nu \beta \alpha \lambda} \quad ,
\end{equation}
where $b_\alpha$ and $d_{\alpha \lambda}$ are coefficients that carry
the information on the transport coefficients
\cite{RelaBoltEqua}. Note that we need to recover terms up to the
fifth order of the equilibrium distribution. In principle, this could
be done by the procedure described on this paper, but the resulting
value for $R$ could be unpractically large. Nevertheless, at low
velocities, $U^\mu \sim (1, 0, 0, 0)$, the Maxwell-J\"uttner
distribution can be approximated by the weight function $w(p^0)$, and
in analogy to the discrete case, by $w_i$, and the fourth and fifth
order are recovered via Eq.~\eqref{eqs:eqs}.  As a result, at
relatively low velocities, we expect the non-equilibrium third order
tensor to be also fulfilled.  Therefore, the transport coefficients
for an ultrarelativistic gas, i.e. $\mu = 0$ for the bulk viscosity,
$\eta = (2/3) P \tau$ for the shear viscosity, and $\lambda = (4/5 T)
P \tau$ for the thermal conductivity, also apply to our model.

To discretize the relativistic Boltzmann equation, we first implement
the time transformation described in the previous section and
integrate in time Eq.~\eqref{Boltzmann:eq2} between $t'$ and $t' +
\delta t'$.  This yields:
\begin{equation}
  f ( x^a + \vartheta^a \delta t', t' + \delta t') - f (x^a, t') =
  -\frac{p^{\mu} U_\mu}{ \tau' p^0}( f - f^{\rm eq} ) \delta t' \quad .
\end{equation}
By changing $p^\mu \rightarrow p_i^\mu$, $f \rightarrow f_i$ and
$\vartheta^a \rightarrow \vartheta_i^a$, we obtain
\begin{equation}\label{rlb:disc}
  f_i ( x^a + \vartheta_i^a \delta t', t' + \delta t') - f_i (x^a, t') =
  -\frac{p_i^{\mu} U_\mu}{ \tau' p_i^0}( f_i - f_i^{\rm eq} ) \delta t' \quad .
\end{equation}
This relativistic lattice Boltzmann equation presents an exact
streaming at the left hand side, and the collision operator at the
right hand side looks exactly like its continuum version.  Therefore,
the conservation laws for the number of particles density 4-flow, and
the momentum-energy tensor, are also fulfilled, as long as they are
obtained by using the Landau-Lifshitz decomposition.  This means that,
first, we need to calculate the momentum-energy tensor,
\begin{equation}
  T^{\alpha \beta} = \sum_{i=1}^{128} f_i p_i^\alpha p_i^\beta \quad ,
\end{equation}
and with this tensor, we solve the eigenvalue problem,
\begin{equation}
  T^{\alpha \beta} U_\beta = T_E^{\alpha \beta} U_\beta = \epsilon U^\alpha \quad ,
\end{equation}
obtaining the energy density $\epsilon$ and the 4-vectors
$U^\alpha$.
Subsequently, the particle density can be calculated by
\begin{equation}\label{densitycal}
  n = U_\mu N_E^\mu = U_\mu N^\mu = \sum_{i=1}^{128} f_i p_i^\mu U_\mu \quad .
\end{equation}
The temperature $T$ is obtained by using the equation of state for the
ultrarelativistic gas, $\epsilon = 3 n T$. The transport coefficients
are the same as in the continuum case, with the lattice correction
resulting from second order Taylor expansion of the streaming term.
All factored in, the coefficients take the following expression $\mu =
0$, $\eta = (2/3) P (\tau' - \delta t'/2)$, and $\lambda = (4/5 T) P
(\tau' - \delta t'/2)$. Note that reverting back the time
transformation, we can write the transport coefficients as $\eta =
(2/3) P (\tau - \delta t/2)/R$, and $\lambda = (4/5 T) P (\tau -
\delta t/2)/R$.

Summarizing, the present model does not present spurious dissipation
in the number of particle conservation equation, in contrast to
previous RLB schemes \cite{rlbPRL, rlbPRD, rlbhupp}, and also improves
the dissipative terms given by the multi-relaxation time scheme
\cite{mrtrlbPRD}. In addition, it realizes the expansion of the
Maxwell-J\"uttner distribution on a cubic lattice, in contrast to
Ref.~\cite{PRCpaul}.  We can also construct a relativistic lattice
Boltzmann model that recovers only up to second order (momentum-energy
tensor), to compare with the third order model and determine the
influence of the third order moment in the expansion. Details of the
second order model can be found in Appendix \ref{second:order}.

\section{Numerical Validation}\label{sec:num}
\begin{figure}
\centering
\includegraphics[scale=0.35]{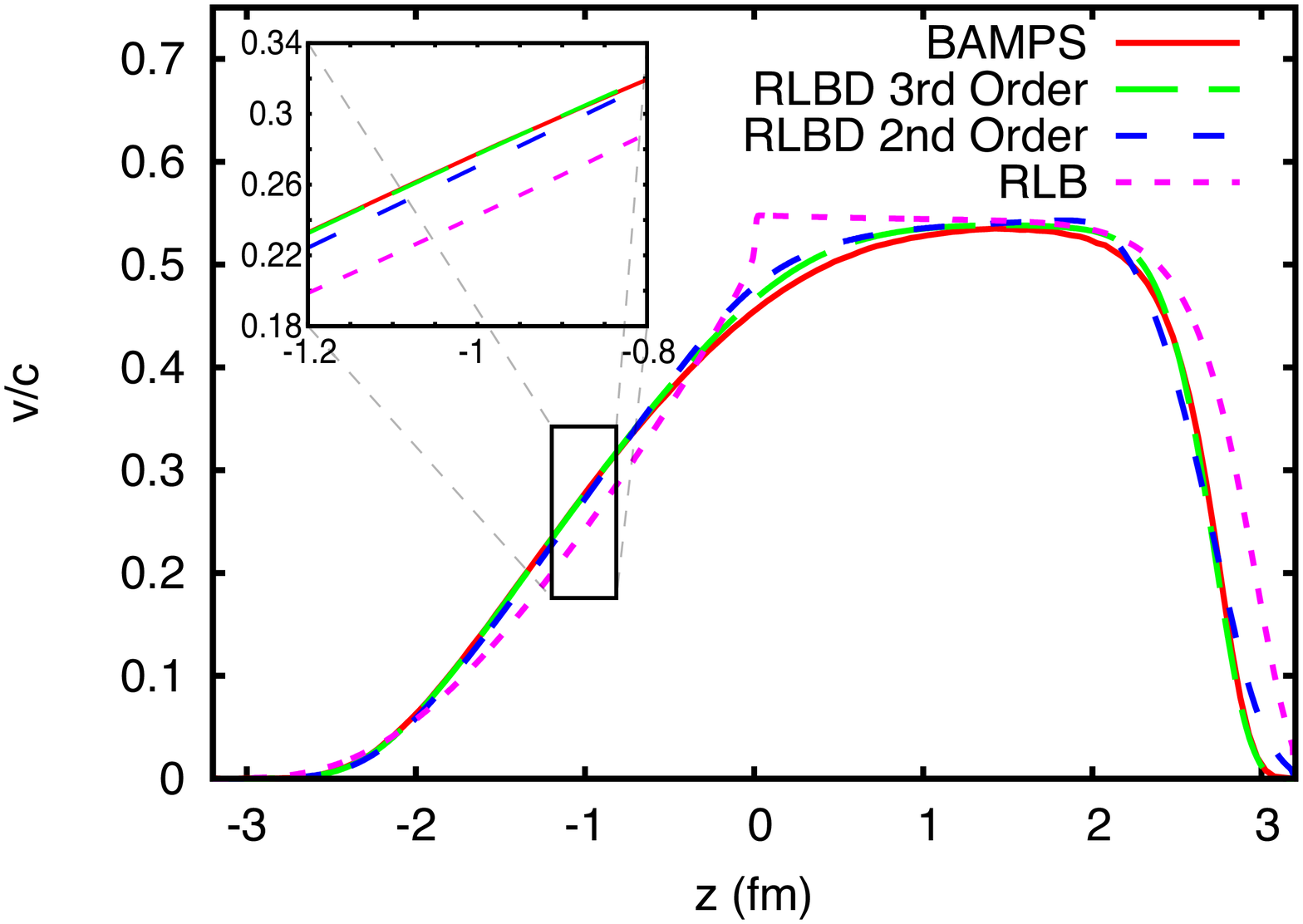}
\includegraphics[scale=0.35]{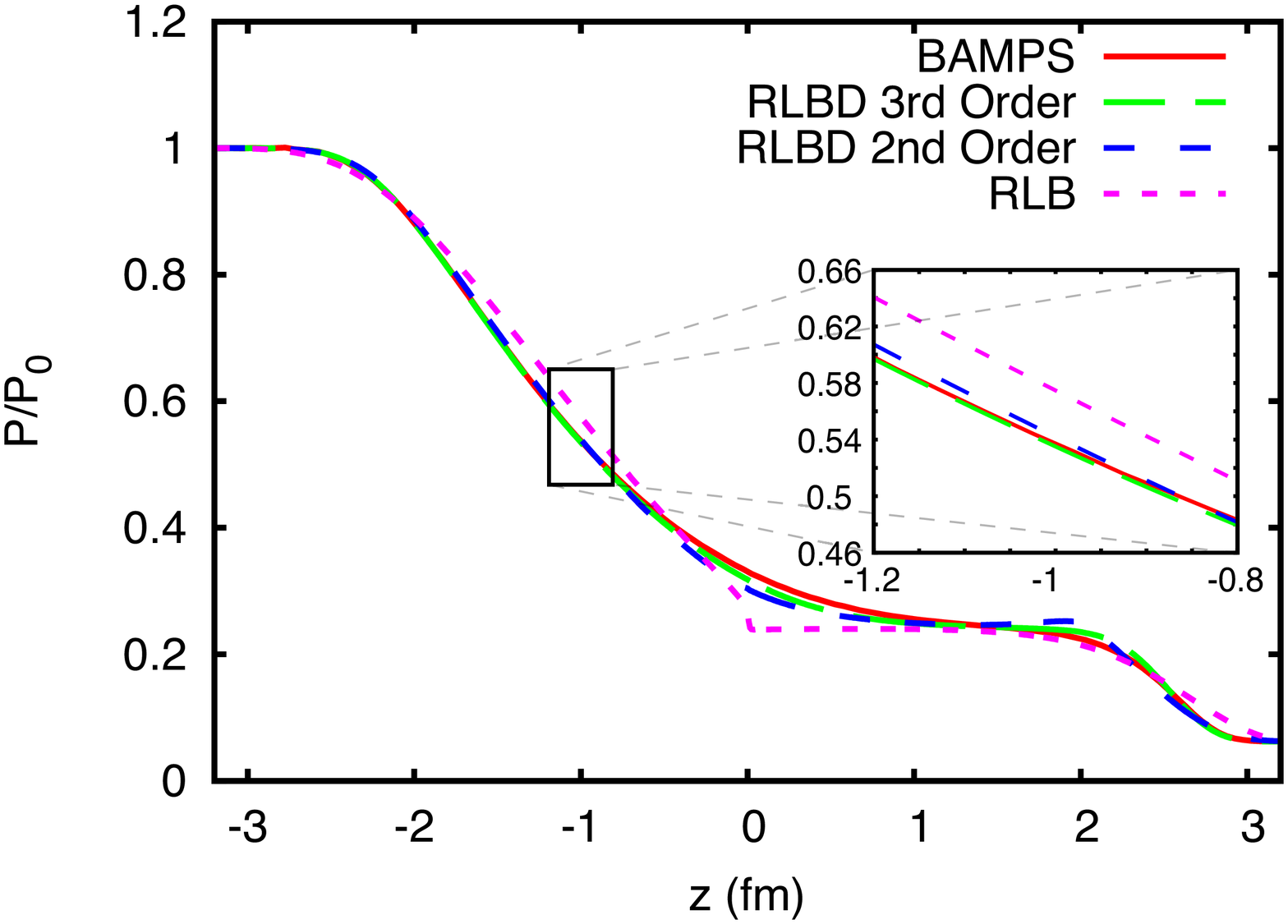}
\caption{Velocity (top) and pressure (bottom) profiles as function of
  the $z$-coordinate for the case of a shockwave in quark-gluon
  plasma, with $\eta/s = 0.1$.}
\label{fig4}
\end{figure}
\begin{figure}
\centering
\includegraphics[scale=0.35]{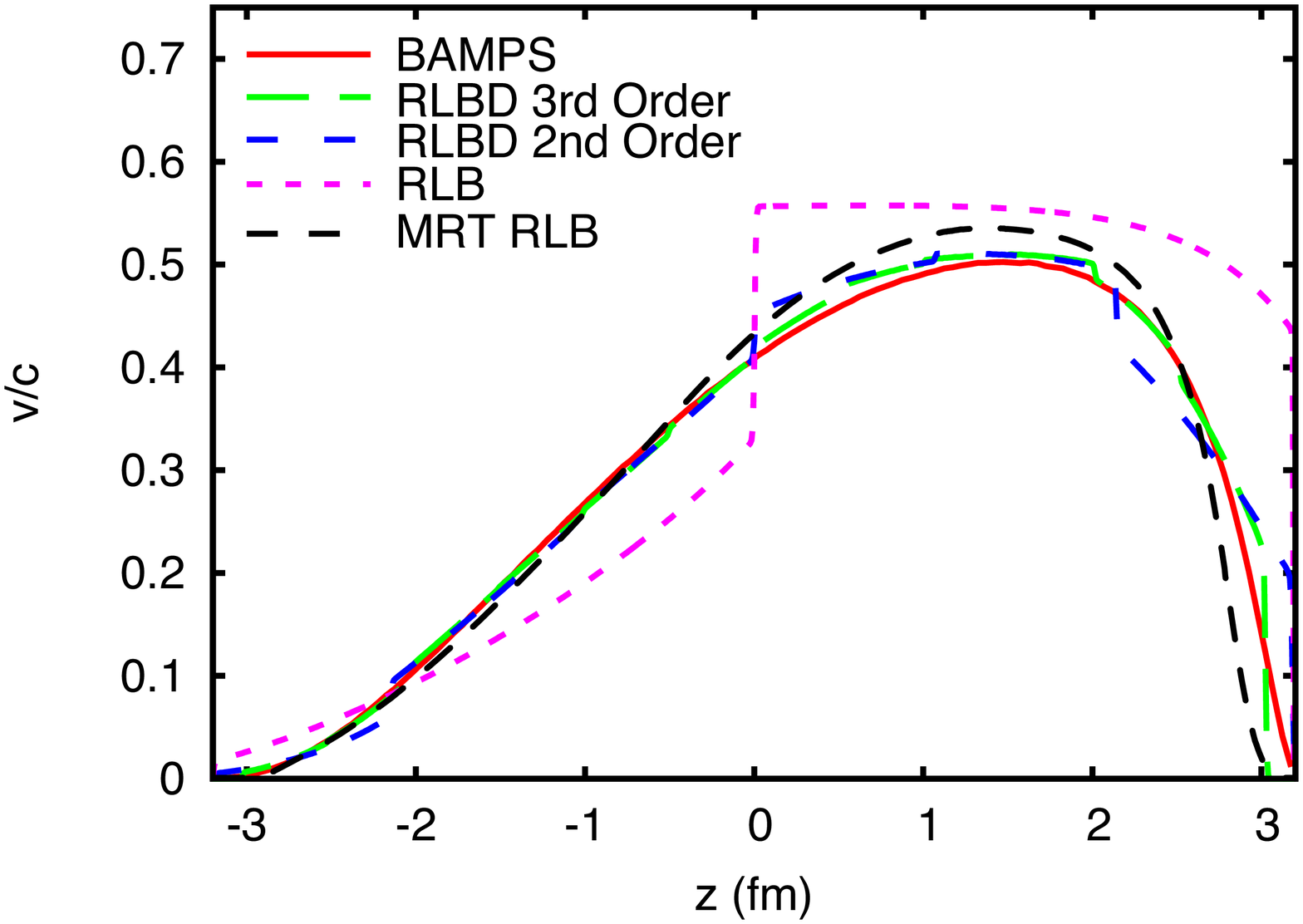}
\includegraphics[scale=0.35]{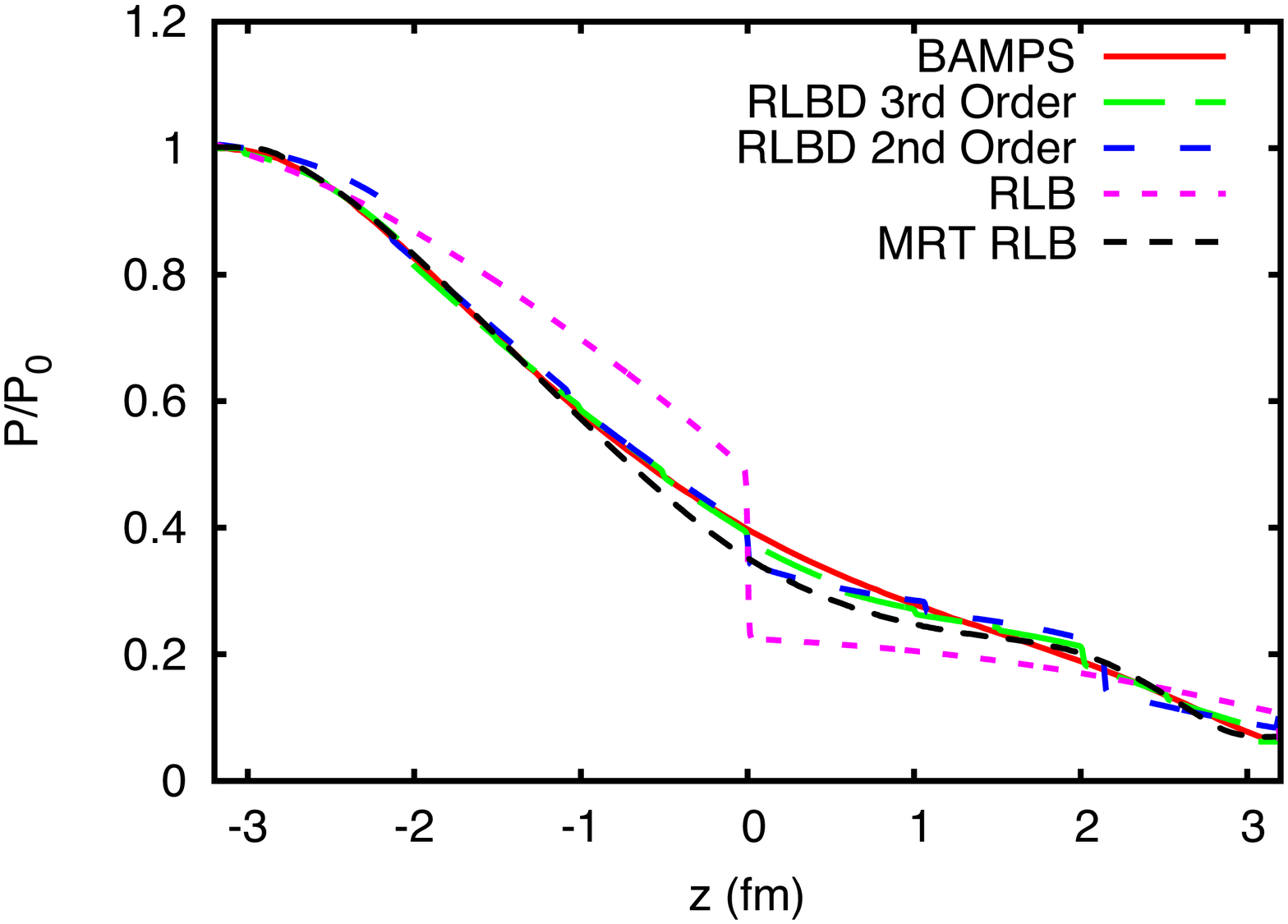}
\caption{Velocity (top) and pressure (bottom) profiles as function of
  the $z$-coordinate for the case of a shockwave in quark-gluon
  plasma, with $\eta/s = 0.5$.}
\label{fig5}
\end{figure}

In order to validate our model, we solve the Riemann problem for a
quark-gluon plasma and compare the results with BAMPS and two previous
relativistic Boltzmann models. The first one, proposed by Mendoza et
al. \cite{rlbPRL, rlbPRD} and later improved by Hupp et al.
\cite{rlbhupp}, which we will denote simply by RLB, and the second
one, which is a recent extension of the RLB developed by Li et al.
\cite{mrtrlbPRD} to include multi-relaxation time, which we will
denote by MRT RLB.  BAMPS was developed by Xu and Greiner \cite{BAMPS}
and applied to the Riemann problem in quark-gluon plasma by Bouras et
al. \cite{BAMPSs}. Since BAMPS solves the full relativistic Boltzmann
equation, we take its result as a reference to access the accuracy of
our model. However, we keep in mind that BAMPS also produces
approximate solutions. The present model is hereafter denoted by RLBD
(RLB with Dissipation).

For small ratios $\eta/s$, where $s$ is the entropy density, RLB and
MRT RLB reproduced BAMPS results to a satisfactory degree of accuracy.
However, for higher $\eta/s \geq 0.1$ and moderately fast fluids,
$\gamma \sim 1.3$, RLB failed to reproduce the velocity and pressure
profiles \cite{rlbhupp}.  MRT RLB yielded good agreement with the
results at $\eta/s = 0.1$, but presented notable discrepancies for
$\eta/s = 0.5$.  The failure of both RLB and MRT RLB to solve the
Riemann problem for high viscous fluids can be ascribed to their
inability to recover the third order moment of the distribution
\cite{rlbhupp, mrtrlbPRD}.
% \begin{figure}
% \centering
% \includegraphics[scale=0.32]{fig10.eps}
% \includegraphics[scale=0.32]{fig11.eps}
% \caption{Velocity (top) and pressure (bottom) profiles as function of
%   the $z$-coordinate for the case of a shockwave in quark-gluon
%   plasma, with $\eta/s = 0.5$. We compare with the recent developed
%   MRT RLB from Ref.~\cite{mrtrlbPRD}.}
% \label{fig6}
% \end{figure}

In this section, we will study the case of high $\eta/s \geq 0.1$ in a
regime of moderate velocities.  We perform the simulations on a
lattice with $1\times1\times1600$ cells, only half of which are
represented in our domain owing to symmetry condition (the other half
is a mirror, in order to use periodic boundary conditions for
simplicity).  Therefore, our simulation consists of $1\times 1\times
800$ lattice sites, with $\delta x = 0.008fm$ and $\delta t =
\sqrt{41}\; 0.008 fm/c$ for RLBD third order, and $\delta t = 0.024
fm/c$ for RLBD second order.

The initial conditions for the pressure are $P_0=5.43\frac{GeV}{fm^3}$
and $P_1=0.339\frac{GeV}{fm^3}$. In numerical units, they correspond
to $1.0$ and $0.062$, respectively. The initial temperature $z \ge 0$
is $T_1 = 200 MeV$ (in numerical units $0.5$), and $T_0 = 400 MeV$ for
$z < 0$, which corresponds to $1.0$ in numerical units. The entropy
density $s$ is calculated according to the relation, $s = 4 n - n
\ln(n/n^{\rm eq})$, where $n^{\rm eq}$ is the density calculated with
the equilibrium distribution, $n^{\rm eq} = d_G T^3/\pi^2$, with $d_G
= 16$ being the degeneracy of the gluons.

The velocity and pressure profiles at $t = 3.2\frac{fm}c$ with
viscosity-entropy density ratios of $\eta/s = 0.1$, are shown in
Fig.~\ref{fig4}. In this figure, we compare the results with BAMPS and
RLB, where we can see that RLB presents a discontinuity at $z=0$,
while both second order and third order RLBD get closer to the BAMPS
solution.  Since the only difference between second and third order
RLBD is the third order moment of the distribution, we conclude that
at relatively low $\eta/s$, the third order does not play a crucial
role neither in the conservative dynamics nor dissipative dynamics of
the system.  However, note that at $z \sim 3 fm$, the third order
model provides an outstanding fit of the numerical results by BAMPS.

\begin{figure}
\centering
\includegraphics[scale=0.35]{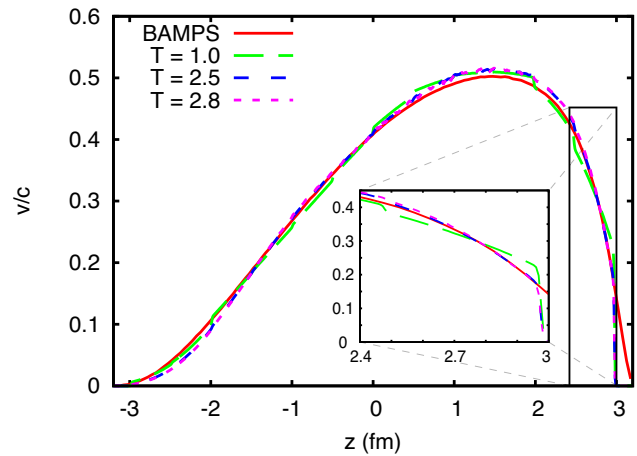}
\caption{Velocity profile as function of the $z$-coordinate for the
  case of a shockwave in quark-gluon plasma, with $\eta/s = 0.5$, by
  increasing the reference numerical temperature in the lattice,
  leading to a smaller relaxation time $\tau$.}
\label{fig7}
\end{figure}

On the other hand, by increasing the ratio $\eta/s$, we see from
Fig.~\ref{fig5} that, while RLB gets worse and the second order RLBD
fixes the discrepancy only in part, the 3rd order RLBD improves
significantly the accuracy of the velocity and pressure profiles.

In Fig.~\ref{fig5}, we also compare the results obtained with MRT RLB
and BAMPS, for $\eta/s = 0.5$. Here, we see that there is again an
improvement, including the attainment of the right value of the
maximum velocity (at $z \sim 1.5 fm$).  In the pressure profile, RLBD
gets closer to BAMPS than MRT RLB in the region of the discontinuity
in the initial condition ($z \sim 0$).

Note that there is a staircase shape in the results of RLBD for
$\eta/s = 0.5$ in Figs.~\ref{fig5}. This is due to the large values
taken by the single relaxation time in order to achieve such shear
viscosity-entropy density ratios, $\tau \sim 20-40$ (in numerical
units), which is beyond the hydrodynamic approximation and therefore
higher order moments (fourth and higher orders) of the distribution
function would be required, which is not fulfilled in our RLBD
model. In order to prove this statement, we have performed separate
simulations, see Fig.~\ref{fig7}, where we observe that by increasing
the value of the reference temperature of the lattice (typically set
at $T = 1$), so as to achieve the same shear viscosity, $\eta = (2/3)
n T (\tau - 1/2 )/R$, the value of $\tau$ decreases and the staircase
disappears. In particular, for $T \ge 2.5$, the results get closer to
the ones with BAMPS, and become independent of the reference
temperature.  Unfortunately, due the discretization procedure used to
develop this model, whenever the reference temperature $T > 4$ the
model becomes unstable, mostly likely because the expanded equilibrium
distribution function takes negative values.

\section{Conclusions and Discussions}\label{sec:diss}

We have introduced a new relativistic lattice Boltzmann model with
improved dissipation, as compared to RLB and MRT RLB.  To this
purpose, we have performed an expansion of the Maxwell-J\"uttner
distribution onto an orthonormal basis of polynomials in the
4-momentum space. In addition, in order to make the model compatible
with a regular cubic lattice, we have performed the expansion in
cartesian coordinates and applied a time transformation, such that
particles travel just the distance necessary to reach lattice nodes,
always at the speed of light. The time transformation generates a
sphere of radius $R$ which intersects the cubic lattice, the
intersection points being lattice nodes by construction.  In addition,
we have reproduced up to second order moment of the equilibrium
distribution, and up to third order moment, finding $R=3$ and
$R=\sqrt{41}$ for second and third order moment compatibility,
respectively.

The discrete energy component of the 4-momentum, $p^0$, has been
calculated by using Gaussian quadrature, the nodes corresponding to
the zeros of the next order polynomial. With this configuration, we
need $90$ vectors for recovering second order and $384$ for the third
order moment case.  However, only $66$ and $128$, respectively, are
actually needed to calculate the moments correctly.

In order to validate the model, we have compared our results with
BAMPS, as well as previous RLB models.  We have found that for $\eta/s
= 0.1$, our model accurately describes the Riemann problem in
quark-gluon plasma, including the expansion up to second order.
However, for the case of $\eta/s = 0.5$, the second order model,
although better than RLB, is less accurate than both MRT RLB and the
third order model.  The third order model yields better results than
the previous RLB, but it develops a staircase shape as a consequence
of the large value of the single relaxation time, which lies beyond
the hydrodynamic regime. We have shown that the staircase pathology
can be tamed by increasing the reference temperature in the model.
Nevertheless, increasing the reference temperature beyond $T=4$ hits
against stability limits of the model.

We may envisage that a multi-relaxation time extension of the present
model would further improve the accuracy of the results.  A similar
improvement may be anticipated by implementing higher order expansions
of the equilibrium distribution.  However, since the transport
coefficients depend on the collision operator, their calculation
within a multi-relaxation time model becomes increasingly involved.
On the other hand, by performing expansions to include higher order
moments, the value of $R$ might become unpractically large, with
several ensuing discretization issues.  Notwithstanding such potential
difficulties, these extensions are surely worth being analyzed in
depth for the future.

\begin{acknowledgments}
  We acknowledge financial support from the European Research Council
  (ERC) Advanced Grant 319968-FlowCCS. Work of I.V.K. was supported by
  the ERC Advanced Grant 291094-ELBM.
\end{acknowledgments}

\appendix

\section{Second order relativistic lattice Boltzmann
  model}\label{second:order}

To construct the second order lattice Boltzmann model, we use the
procedure described in this paper. We have obtained that $R=3$
presents enough symmetries to fulfill the conditions in
Eqs.~\eqref{eqs:eqs}, and the velocity vectors $\vec{\vartheta}$ are
given by, $(\pm 3,0,0)$, $(0,\pm 3, 0)$, $(0, 0, \pm 3)$, $(\pm 2, \pm
1, \pm 2)$, $(\pm 1, \pm 2, \pm 2)$, and $(\pm 2, \pm 2, \pm 1)$. The
values for the discrete $p^0$ come from the solution of the equation,
\begin{equation}
  {\cal R}^{(3)} = \frac{1}{12} p^0 (p^0 - 6)^2 - 2 = 0 \quad ,
\end{equation}
instead of ${\cal R}^{(4)}$ for the case of the third order
expansion. This gives the values $p_l^0 \simeq 0.936$, $3.305$, and
$7.759$. The discrete 4-momentum vectors $p_i^\mu$ are constructed
with Eqs.~\eqref{eq:mrecon}, and \eqref{eq:mrecon2}, and they are in
total, ${\cal N} = 3 \times 30 = 90$. However, as in the third order
expansion, we have retained the minimal amount, out of $90$, that are
necessary to recover the second order moment, by imposing the maximum
number of $w_i$ to be zero. This gives only $66$ 4-momentum
vectors. The value of the weight functions for every momentum vector
and the relation with the $30$ directions are given in the
Supplementary Material \cite{supp}.  In Fig.~\ref{fig2} we report the
spatial configuration of the vectors $\vec{\vartheta}_i$.

\begin{figure}
\centering
\includegraphics[scale=0.8]{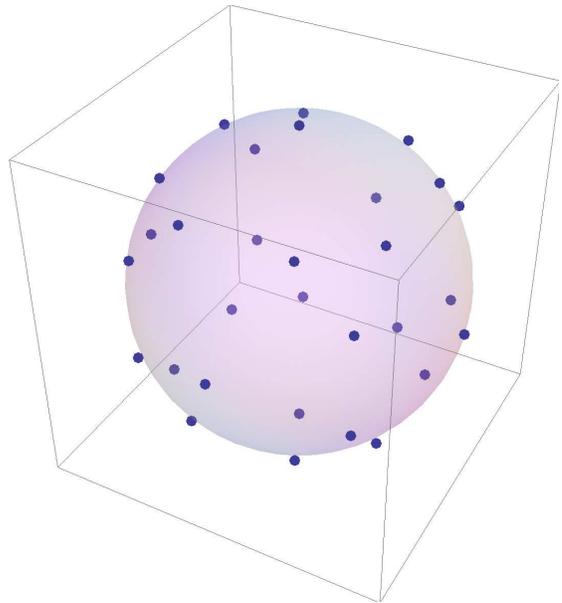}
\caption{Directions of the velocity vectors $\vec{\vartheta}_i$ to
  recover up to the second order moment of the Maxwell-J\"uttner
  distribution, namely the momentum-energy tensor. The radius of the
  sphere is $R=3$. The points represent lattice sites belonging to the
  surface of the sphere.}
\label{fig2}
\end{figure}
The discrete version of the relativistic Boltzmann equation,
Eq.~\eqref{rlb:disc}, still applies and the discrete equilibrium
distribution function is written in detail in Appendix
\ref{eqfuncapp}, Eq.~\eqref{eq:equi2nd}. However, due to the fact that
the third order moment is not satisfied, an analytical theory to
calculate the transport coefficients would be very complicated and
goes beyond the scope of this work. Therefore, we have calculated
numerically only the shear viscosity, by matching the results for low
velocity with the third order moment model. This, in order to compare
the results of both expansions with other models in the
literature. This gives a shear viscosity $\eta_{\rm 2nd} \sim (1/7) P
(\tau - \delta t/2)/R$.  We could, in principle, calculate the third
order moment associated with the equilibrium distribution given by
Eq.~\eqref{eq:equi2nd}, and, by applying the Grad method, compute the
other transport coefficients.  However, this procedure would need to
be performed entirely numerically, since the weights $w_i$ and
4-momentum vectors $p_i^\mu$ are only known numerically.  Since the
main purpose of this paper is to improve the description of
dissipative effects by performing the third order expansion and place
it on a cubic lattice, we are not interested in the bulk viscosity and
the thermal conductivity for this case, and leave this task for future
work.

\section{Equilibrium Distribution Functions}\label{eqfuncapp}

The equilibrium distribution function capable to recover the first and
second order moments of the equilibrium distribution is calculated by
using up to the second order polynomials in Eq.~\eqref{eq:expanded},
namely the $14$ polynomials $J_k$ with $k=0, ..., 13$, obtaining
\begin{widetext}
  \begin{equation}\label{eq:equi2nd}
  \begin{aligned}
    f_i^{\rm eq} &= \frac{n w_i}{4 T} \bigg [ {p_i^0}^2 \left(T^2
      \left(2 {U^0}^2-{U^x}^2-{U^y}^2-1\right)-2 T {U^0}+1\right)+2
    {p_i^0} (T (T ({U^0} ({p_i^x} {U^x}+{p_i^y} {U^y}+{p_i^z}
    {U^z}-4 {U^0})+1) \\ &-{p_i^x} {U^x}-{p_i^y} {U^y}-{p_i^z} {U^z}+7
    {U^0})-4)+T^2 \bigg ({p_i^x}^2 \left(-{U^0}^2+2
      {U^x}^2+{U^y}^2+1\right)+2 {p_i^x} {U^x} ({p_i^y} {U^y}+{p_i^z}
    {U^z}-4 {U^0}) \\ &+{p_i^y}^2 \left(-{U^0}^2+{U^x}^2+2
      {U^y}^2+1\right)+2 {p_i^y} {U^y} ({p_i^z} {U^z}-4 {U^0})+8 {U^0}
    ({U^0}-{p_i^z} {U^z})-2\bigg ) \\ &+2 T (5 ({p_i^x}
    {U^x}+{p_i^y} {U^y}+{p_i^z} {U^z})-8 {U^0})+12\bigg ] \quad ,
  \end{aligned}
\end{equation}
\end{widetext}
For the case of the third order moment expansion, we repeat the same
procedure, using all the polynomials ($k=0, ..., 29$).  This leads to
the following expressions:
\begin{widetext}
\begin{equation}\label{eq:equi3rd}
  \begin{aligned}
    f_i^{\rm eq} &= \frac{n w_i}{12 T} \bigg [{p_i^0}^3 (T
    {U^0}-1) \left(T^2 \left(4 {U^0}^2-3
        \left({U^x}^2+{U^y}^2+1\right)\right)-2 T {U^0}+1\right) \\
    &-{p_i^0}^2 \bigg (T^3 \bigg (-2 {U^0}^2 (3 {p_i^x} {U^x}+3
    {p_i^y} {U^y}+2 {p_i^z} {U^z})+\left({U^x}^2+{U^y}^2+1\right) (3
    {p_i^x} {U^x}+3 {p_i^y}{U^y}+{p_i^z} {U^z}) \\ &+36 {U^0}^3-6
    {U^0} \left(3 {U^x}^2+3 {U^y}^2+4\right)\bigg )+3 T^2 \left(2
      {U^0} ({p_i^x} {U^x}+{p_i^y} {U^y}+{p_i^z} {U^z})-22 {U^0}^2+7
      \left({U^x}^2+{U^y}^2\right)+9\right) \\ &-3 T ({p_i^x}
    {U^x}+{p_i^y} {U^y}+{p_i^z} {U^z}-14 {U^0})-15\bigg )-3 {p_i^0}
    \bigg (T^3 \bigg({U^0}^3
    \left({p_i^x}^2+{p_i^y}^2-24\right)-{U^0} \bigg ({p_i^x}^2 \left(2
      {U^x}^2+{U^y}^2+1\right) \\ &+2 {p_i^x} {U^x} ({p_i^y}
    {U^y}+{p_i^z} {U^z})+{p_i^y} \left({p_i^y} {U^x}^2+2 {p_i^y}
      {U^y}^2+{p_i^y}+2 {p_i^z} {U^y} {U^z}\right)-12\bigg)+12 {U^0}^2
    ({p_i^x} {U^x}+{p_i^y} {U^y}+{p_i^z} {U^z}) \\ &-2 ({p_i^x}
    {U^x}+{p_i^y} {U^y}+{p_i^z} {U^z})\bigg)+T^2 \bigg ({p_i^x}^2
    \left(-{U^0}^2+2 {U^x}^2+{U^y}^2+1\right)+2 {p_i^x} {U^x} ({p_i^y}
    {U^y}+{p_i^z} {U^z}-11 {U^0}) \\ &+{p_i^y}^2
    \left(-{U^0}^2+{U^x}^2+2 {U^y}^2+1\right)+2 {p_i^y} {U^y} ({p_i^z}
    {U^z}-11 {U^0})-22 {p_i^z} {U^0} {U^z}+56 {U^0}^2-14\bigg ) \\ &+2
    T (6 ({p_i^x} {U^x}+{p_i^y} {U^y}+{p_i^z} {U^z})-25
    {U^0})+20\bigg)+T \bigg ({p_i^x}^3 T^2 {U^x} \left(-3
      {U^0}^2+4 {U^x}^2+3 {U^y}^2+3\right) \\ &+{p_i^x}^2 T \left(3
      \left({U^0}^2-2 {U^x}^2-{U^y}^2-1\right) (-{p_i^y} T {U^y}+6
      T {U^0}-7)+{p_i^z} T {U^z} \left(-{U^0}^2+4
        {U^x}^2+{U^y}^2+1\right)\right) \\ &+3 {p_i^x} {U^x} \bigg(T
    \bigg (T \left({p_i^y}^2 \left(-{U^0}^2+{U^x}^2+2
        {U^y}^2+1\right)+2 {p_i^y} {U^y} ({p_i^z} {U^z}-6 {U^0})-12
      {p_i^z} {U^0} {U^z}+24 {U^0}^2-4\right) \\ &+14 {p_i^y} {U^y}+14
    {p_i^z} {U^z}-48 {U^0}\bigg )+30\bigg)+{p_i^y}^3 T^2 {U^y}
    \left(-3 {U^0}^2+3 {U^x}^2+4 {U^y}^2+3\right) \\ &+{p_i^y}^2 T
    \left({p_i^z} T {U^z} \left(-{U^0}^2+{U^x}^2+4
        {U^y}^2+1\right)+3 (6 T {U^0}-7) \left({U^0}^2-{U^x}^2-2
        {U^y}^2-1\right)\right) \\ &+6 {p_i^y} {U^y} \left(T \left(2
        T \left(-3 {p_i^z} {U^0} {U^z}+6 {U^0}^2-1\right)+7 {p_i^z}
        {U^z}-24 {U^0}\right)+15\right)+6 {p_i^z} {U^z} \left(2 T^2
      \left(6 {U^0}^2-1\right)-24 T {U^0}+15\right) \\ &-24 {U^0}
    \left(T^2 \left(2 {U^0}^2-1\right)-5 T {U^0}+5\right)\bigg
    )-30 \left(T^2-2\right)\bigg ] \quad .
  \end{aligned}
\end{equation}
\end{widetext}

\bibliography{report}

\end{document}